\documentclass[journal,10pt,twoside]{IEEEtran}
\normalsize
\newif\ifonecolumn
\IEEEoverridecommandlockouts
\usepackage[final]{graphicx}
\usepackage{times}
\usepackage{amsfonts}
\usepackage{epsfig}
\usepackage{amssymb}
\usepackage{amsmath}
\usepackage{algorithm}
\usepackage{algorithmic}
\usepackage{citesort}
\usepackage{psfrag}
\usepackage{balance}
\usepackage{comment}
\usepackage{color}

\newcommand{\beq}{\begin{equation}}
\newcommand{\eeq}{\end{equation}}
\newcommand{\ben}{\begin{enumerate}}
\newcommand{\een}{\end{enumerate}}
\newcommand{\bit}{\begin{itemize}}
\newcommand{\eit}{\end{itemize}}

\newtheorem{example}{Example}
\newtheorem{lemma}{Lemma}

\newtheorem{theorem}{Theorem}

\newcounter{mytempeqncnt}

\begin{document}
\title{On the Minimum/Stopping Distance of Array Low-Density Parity-Check Codes}

\author{Eirik~Rosnes,~\IEEEmembership{Senior Member,~IEEE}, 
        Marcel A.\ Ambroze,~\IEEEmembership{Member,~IEEE},\\and~Martin~Tomlinson,~\IEEEmembership{Senior Member,~IEEE}

\thanks{The work of E.\ Rosnes  was supported by the Research Council of Norway (NFR) under grant 183316 and by Simula@UiB. The material in this paper was presented in part at the 2014 IEEE International Symposium on Information Theory, Honolulu, HI, June/July 2014.}%
\thanks{E.\ Rosnes was with Ceragon Networks AS, Kokstadveien 23, N-5257 Kokstad, Norway. He is now with the Selmer Center, Department of Informatics, University of Bergen, N-5020 Bergen, Norway, and the Simula Research Lab. E-mail: eirik@ii.uib.no.}%
\thanks{M.\ A.\ Ambroze and M.\ Tomlinson are with the Fixed and Mobile Communications Research Center, University of Plymouth, PL4 8AA, UK. E-mail: \{marcel.ambroze, martin.tomlinson\}@plymouth.ac.uk.}}

\maketitle

%
%
%

\maketitle

\begin{abstract}
In this work, we study the minimum/stopping distance of array low-density parity-check (LDPC) codes. 
An array LDPC code is a quasi-cyclic LDPC code specified by two integers $q$ and $m$, where $q$ is an odd prime  and $m \leq q$. In the literature, the minimum/stopping distance of these codes (denoted by  $d(q,m)$ and $h(q,m)$, respectively) has been thoroughly studied for $m \leq 5$. Both exact results, for small values of $q$ and $m$, and general (i.e., independent of $q$) bounds have been established. 
For $m=6$, the best known minimum distance upper bound, derived by Mittelholzer (\emph{IEEE Int.\ Symp.\ Inf.\ Theory}, Jun./Jul.\ 2002), is $d(q,6) \leq 32$. 
In this work, we derive an improved upper bound of $d(q,6) \leq 20$ and a new upper bound $d(q,7) \leq 24$ by using the concept of a \emph{template support matrix} of a codeword/stopping set. 
The bounds are tight with high probability in the sense that we have not been able to find codewords of strictly lower weight  for several values of $q$ using a minimum distance probabilistic algorithm.  
Finally, we provide new specific minimum/stopping distance results for $m \leq 7$ and low-to-moderate values of $q \leq 79$.
\end{abstract}


\section{Introduction}

In this paper, we consider the array low-density parity-check (LDPC) codes, originally introduced by Fan in \cite{fan00},  and their minimum/stopping distance.  Array LDPC codes are specified by two integers $q$ and $m$, where $q$ is an odd prime and $m \leq q$. Furthermore, in this work, $\mathcal{C}(q,m)$ will denote the array LDPC code with parameters $q$ and $m$, and $d(q,m)$ (respectively $h(q,m)$)  its minimum (respectively stopping) distance. 


Since the original work by Fan, several authors have considered the \emph{structural} properties of these codes (see, e.g., \cite{mit02,yan03,sug08,esm09,esm11,liu10,dol10}). For high rate and
moderate length, these codes perform well under iterative decoding, and they are also well-suited for practical implementation due to their regular structure \cite{olc03,bha05}.


The minimum distance of these codes was first analyzed by Mittelholzer in \cite{mit02}, where general (i.e., independent of $q$) minimum distance upper bounds for $m \leq 6$ were provided. Subsequently, Yang and Helleseth \cite{yan03} investigated the minimum
distance of these codes in an algebraic way by first proving that the codes are invariant under a doubly transitive group of ``affine'' permutations. Then, they proved the general lower bound $d(q,4) \geq 10$, for $q > 7$, on the minimum distance. In \cite{sug08}, the general upper bounds $d(q,4) \leq 10$  and $d(q,5) \leq 12$ 
on the minimum distance were proved. Furthermore, by combining these bounds with the results in \cite{yan03}, it follows that $d(q,4)=10$  and that $d(q,5)$ is either $10$ or $12$, for $q > 7$. In summary,
%
%
\begin{displaymath}
d(q,m) \leq \begin{cases}
6, & \text{if $m=3$, with equality for $q \geq 5$ \cite{yan03}}\\
10, & \text{if $m=4$, with equality for $q > 7$ \cite{yan03,sug08}}\\
12, & \text{if $m=5$, with exact value either} \\
& \text{$10$ or $12$ for $q > 7$ \cite{yan03,sug08}}\\
32, & \text{if $m=6$ \cite{mit02}.} \end{cases}
\end{displaymath}

The case $m=6$ has not been treated in the literature before, except for in the initial work of Mittelholzer \cite{mit02}. In this work, we will consider this case in more detail as well as the case $m=7$, both from an experimental point of view and by deriving an improved upper bound on $d(q,6)$ and a new upper bound on $d(q,7)$. 

This paper is organized as follows. 
In Section~\ref{sec:prelim}, some of the basic notation is introduced and the definition of array LDPC codes is given.
The concept of a \emph{template support matrix} is also introduced.
In Section~\ref{sec:algo}, an heuristic is presented that will be used to infer a \emph{candidate} template support matrix.
The heuristic analyzes the \emph{graphical cycle structure} of support matrices of codewords/stopping sets for different values of $q$, with $m$ fixed.
In Section~\ref{sec:bound}, we use the (candidate) template support matrix found in Section~\ref{sec:algo} to formally prove the improved upper bound $d(q,6) \leq 20$.
Furthermore, in Section~\ref{sec:bound_m7}, we present a template support matrix for $m=7$, found by using the heuristic of Section~\ref{sec:algo}, which is used to formally prove the new upper bound $d(q,7)\leq24$.
In Section~\ref{sec:results}, new minimum/stopping distance results are presented for fixed values of $m \leq 7$ and $q \leq 79$. 
Finally, in Section~\ref{sec:conclu}, we draw the conclusions and present some directions for future work.

\section{Preliminaries} \label{sec:prelim}

The array LDPC code $\mathcal{C}(q,m)$, with parameters $q$ and $m$, has length $q^2$ and can be defined by the parity-check matrix
\begin{equation} \label{eq:pcmatrix}
\mathbf{H}(q,m) = \begin{pmatrix}
\mathbf{I} & \mathbf{I} & \mathbf{I} & \cdots & \mathbf{I} \\
\mathbf{I} & \mathbf{P} & \mathbf{P}^2 & \cdots & \mathbf{P}^{q-1} \\
\mathbf{I} & \mathbf{P}^2 & \mathbf{P}^4 & \cdots & \mathbf{P}^{2(q-1)} \\
& & \vdots  & & \vdots \\
\mathbf{I} & \mathbf{P}^{m-1} & \mathbf{P}^{2(m-1)} & \cdots & \mathbf{P}^{(m-1)(q-1)} \end{pmatrix}
\end{equation}
where $\mathbf{I}$ is the $q \times q$ identity matrix and $\mathbf{P}$  is a $q \times q$ permutation matrix defined by
\begin{displaymath}
\mathbf{P} = \begin{pmatrix}
0 & 0 & \cdots & 0 & 1 \\
1 & 0 & \cdots & 0 & 0 \\
0 & 1 & \cdots & 0 & 0 \\
& \vdots & &  \vdots & \\
0 & 0 & \cdots & 1 & 0 \end{pmatrix}.
\end{displaymath}
Since the number of ones in each row of the matrix in (\ref{eq:pcmatrix}) is $q$ and the number of ones in each column is $m$, the array LDPC codes are $(m,q)$-regular codes. Furthermore, it is not hard to see that the parity-check matrix in (\ref{eq:pcmatrix}) has rank $qm-m+1$, from which it follows that the dimension of $\mathcal{C}(q,m)$ is $q^2-qm+m-1$. 

In \cite{yan03}, a new representation for $\mathbf{H}(q,m)$ was introduced. In particular, since each column of the parity-check matrix $\mathbf{H}(q,m)$ has $m$ blocks and each block is a permutation of $(1,0,0,\dots,0,0)^T$, where $(\cdot)^T$ denotes the transpose of its argument, we can represent each column as a vector of integers between $0$ and $q-1$, where
\begin{equation} \label{eq:rep}
i \triangleq \left(\overbrace{0,\dots,0}^{i},1,\overbrace{0,\dots,0}^{q-i-1} \right)^T
\end{equation}
i.e., the $1$-positions are associated with the integers modulo $q$. 
 Furthermore, it follows from (\ref{eq:pcmatrix}) and the integer representation in (\ref{eq:rep})  that any column in an array LDPC code parity-check matrix is of the form 
\begin{equation} \label{eq:form}
(x,x+y,x+2y,\dots,x+(m-1)y)^T \pmod q
\end{equation}
where $x$ and $y$ are integers between $0$ and $q-1$. Thus, a column can be specified by two integers $x$ and $y$. Also, note that since there are $q^2$ distinct columns in an array LDPC code parity-check matrix, any pair $(x,y) \in \mathbb{Z}_q^2$ where $\mathbb{Z}_q=\{0,\dots,q-1\}$ specifies a valid column.

In the following, the \emph{support matrix} of a codeword/stopping set will be the submatrix of $\mathbf{H}(q,m)$ corresponding to the \emph{support set} of the codeword/stopping set, i.e., we keep the columns of  $\mathbf{H}(q,m)$ whose column indices coincide with the support set of the codeword/stopping set. Also, we will use the integer representation in (\ref{eq:rep}) for the columns of the submatrix.

Furthermore, a \emph{template} support matrix with parameters $m$, $q$, $w$, and $q_0$  is formally defined as an $m \times w$ matrix with entries that are functions of $q$ and such that it is the support matrix (possibly column-permuted) of a codeword/stopping set of weight/size $w$ of  $\mathcal{C}(q,m)$ for all $q \geq q_0$. 
The specific matrix which results when a template support matrix is evaluated for a specific value of $q$ is called an \emph{instance} of the template support matrix.

\section{Deriving Upper Bounds on $d(q,m)$} \label{sec:algo}

In this section, we describe an heuristic which can be used to derive upper bounds on the minimum/stopping distance of array LDPC codes. For simplicity, we will only consider the codeword case (the stopping set case is similar and is explicitly considered in Section~\ref{sec:stopping} below). 
The heuristic is a three-step procedure:
\begin{enumerate}
\item In the first step, pairs of codewords $\mathbf{c}_1 \in \mathcal{C}(q_1,m)$ and $\mathbf{c}_2 \in \mathcal{C}(q_2,m)$, $q_1 < q_2$ and $m$ fixed, where $\mathbf{c}_1$ and $\mathbf{c}_2$ have the same \emph{graphical cycle structure} (a concept to be defined later), are identified. 
\item The second step is to infer a \emph{candidate} template support matrix (which may or may not exist) such that the instances for $q=q_1$ and $q=q_2$ are the support matrices (possibly column-permuted) of the two codewords $\mathbf{c}_1$ and $\mathbf{c}_2$, respectively. 
We emphasize here that the inferred matrix is only a \emph{candidate} template support matrix, since a formal proof is needed to show that all instances for $q \geq q_0$, for some $q_0$, are in fact valid (possibly column-permuted) support matrices. 
\item The third step is a formal proof that the instances of the candidate template support matrix are indeed valid  
(possibly column-permuted)  
support matrices of codewords for all possible values of $q$ larger than or equal to $q_0$. 
\end{enumerate}

One way to find an upper bound on the minimum/stopping distance for a fixed value of $m$ which is also independent of $q$ (if such a bound exists), is to identify a common \emph{structure} of codewords/stopping sets for different values of $q$. This justifies the first step of the heuristic above which looks for a common underlying structure to the pairs of codewords $\mathbf{c}_1 \in \mathcal{C}(q_1,m)$ and $\mathbf{c}_2 \in \mathcal{C}(q_2,m)$. Then, in the second step, such a common structure  in the form of a template support matrix (valid at least for $q=q_1$ and $q=q_2$) is determined. In the final third step, we try to prove that the candidate template support matrix of the previous step is indeed a valid template support matrix for all $q$ larger than or equal to some threshold value $q_0$.

Finally, we note that all instances of a template support matrix may not have their columns in the order implied by the parity-check matrix in (\ref{eq:pcmatrix}). 
This is obviously not important, since the order of the columns in a support matrix is not relevant (independent of the order, it will represent the same codeword/stopping set).

\subsection{First Step: Graphical Cycle Structure} \label{sec:structure}

Note that for the array LDPC codes there exists a subgroup of the automorphism group which is doubly transitive \cite{yan03}. For convenience of the reader we state the formal result below as a lemma. For details and its proof, we refer the interested reader to \cite[Lemma~2]{yan03}.

Let $\mathcal{T}$ be defined as the set of columns of $\mathbf{H}(q,m)$ using the representation in (\ref{eq:form}), i.e., 
\begin{displaymath}
\mathcal{T} = \left\{(x,x+y,x+2y,\dots,x+(m-1)y)^T{:}\ x,y \in \mathbb{Z}_q \right\}
\end{displaymath}
where the operations are taken modulo $q$.

\begin{lemma} \label{lab:lemma_subgroup}
The array LDPC code $\mathcal{C}(q,m)$ is invariant under the doubly transitive
group of ``affine'' permutations $\Psi$ of the form
\begin{displaymath}
  \begin{array}{rrcl}
\Psi: & \mathcal{T} & \to & \mathcal{T} \\
 & \mathbf{x} & \mapsto &  a \mathbf{x} + \mathbf{b}
\end{array}
\end{displaymath}
where $a \in \mathbb{Z}_q \setminus \{0\}$, $\mathbf{b} \in \mathcal{T}$, and all operations are taken componentwise modulo $q$. 
\end{lemma}


From Lemma~\ref{lab:lemma_subgroup}, it follows that for any codeword $\mathbf{c} \in \mathcal{C}(q,m)$  and coordinates $p_1$ and $p_2$, $0 \leq p_1 < p_2 < q^2$, there exists a codeword $\rho(\mathbf{c})$ (obtained by permuting the coordinates of $\mathbf{c}$ according to a permutation $\rho$ from this subgroup) having $p_1$ and $p_2$ in its support set. Thus, it is always possible to permute any codeword (using permutations from this subgroup) such that the corresponding support matrix contains the columns $(0,0,0,\dots,0)^T$ and $(q-1,0,1,\dots,q-2)^T$. This is the case since these columns will always be in the parity-check matrix $\mathbf{H}(q,m)$ for all valid values of $q$ and $m$. In particular, the column $(0,0,0,\dots,0)^T$ (respectively $(q-1,0,1,\dots,q-2)^T$) is generated by $x=0$ and $y=0$ (respectively $x=q-1$ and $y=1$) using the representation in (\ref{eq:form}).

As argued above, the support matrix can be regarded as an $m \times w$ matrix of integers modulo $q$, where $w$ is the weight  of the underlying codeword.  From this matrix we can make a bipartite graph, denoted by $G^{(i,j)}=G(V^{(i,j)},E^{(i,j)})$, for each pair of rows $(i,j)$, $i<j$. The vertex set $V^{(i,j)}$ partitions into two distinct sets which we denote by $V^{(i)}$ and $V^{(j)}$, respectively. Now, for each \emph{distinct} entry in the $i$th row of the support matrix we associate a node in the vertex set $V^{(i)}$. Thus, if there are two (or more) identical entries in the $i$th row of the support matrix, then they will correspond to the same vertex in  $V^{(i)}$. Similarly, for each \emph{distinct} entry in the $j$th row of the support matrix we associate a node in the vertex set $V^{(j)}$. Furthermore, there will be an edge from a vertex $v^{(i)} \in V^{(i)}$ to a vertex $v^{(j)} \in V^{(j)}$ if and only if there exists a column in the support matrix in which the entry corresponding to $v^{(i)}$ appears as the $i$th element and the entry corresponding to $v^{(j)}$ appears as the $j$th element. 
In the following, we will refer to the graphs $G^{(i,j)}$ as the  \emph{support matrix graphs}. 
For convenience, we let $v^{(i)}_{\alpha}$ denote the vertex in $V^{(i)}$ representing the entry (or entries) with value $\alpha$ in the $i$th row of the support matrix of a codeword. 
Also,
due to the automorphism group (see Lemma~\ref{lab:lemma_subgroup}),  
we will assume that the support matrix of a codeword  contains the columns $(0,0,0,\dots,0)^T$ and $(q-1,0,1,\dots,q-2)^T$.


Let  $\mathbf{c}_1 \in \mathcal{C}(q_1,m)$ and $\mathbf{c}_2 \in \mathcal{C}(q_2,m)$, $q_2 > q_1$, be two distinct \emph{minimal} codewords of the same Hamming weight, where a \emph{minimal} codeword is a codeword that does not have the support set of a nonzero codeword as a proper subset of its own support set. From each of the corresponding support matrices we build the support matrix graphs $G^{(i,j)}$ for each pair of rows $(i,j)$, $0 \leq i < j < m$, as outlined above. The graphs corresponding to $\mathbf{c}_1$ and $\mathbf{c}_2$ are denoted by  $G_{\mathbf{c}_1}^{(i,j)}$ and $G_{\mathbf{c}_2}^{(i,j)}$, respectively. Now, $\mathbf{c}_1$ and $\mathbf{c}_2$ are said to have the same \emph{graphical cycle structure} (by definition) if and only if the graphs  $G_{\mathbf{c}_1}^{(i,j)}$ and  $G_{\mathbf{c}_2}^{(i,j)}$, for each pair $(i,j)$, 
have the same number of (proper) cycles of a given length containing the edge $(v^{(i)}_{q-1+i \pmod q}, v^{(j)}_{q-1+j \pmod q})$, where $q-1+i \pmod q$ is the $i$th component of the column $(q-1,0,1,\dots,q+m-2)^T$, and also the same number of (proper) cycles of a given length containing the edge $(v^{(i)}_0,v^{(j)}_0)$. In general, in this paper, when speaking about cycles we mean proper cycles, i.e., cycles in which all intermediate nodes are distinct and different from the starting node.




The basic idea is to identify pairs of (minimal) codewords $\mathbf{c}_1 \in \mathcal{C}(q_1,m)$ and $\mathbf{c}_2 \in \mathcal{C}(q_2,m)$, $q_2 > q_1$ and $m$ fixed, with the same graphical cycle structure, since if they do not have the same graphical cycle structure, then it is likely (although not impossible when $q_1$ or $q_2$ is small) 
that their support matrices 
cannot be instances of the same template support matrix. Then, for a pair of (minimal) codewords with the same graphical cycle structure, we would like to infer a template support matrix 
such that the instances for $q=q_1$ and $q=q_2$ are the support matrices (possibly column-permuted) of the codewords $\mathbf{c}_1$ and $\mathbf{c}_2$, respectively.

\begin{figure}[tbp]
  \centerline{\includegraphics[height=0.4\columnwidth]{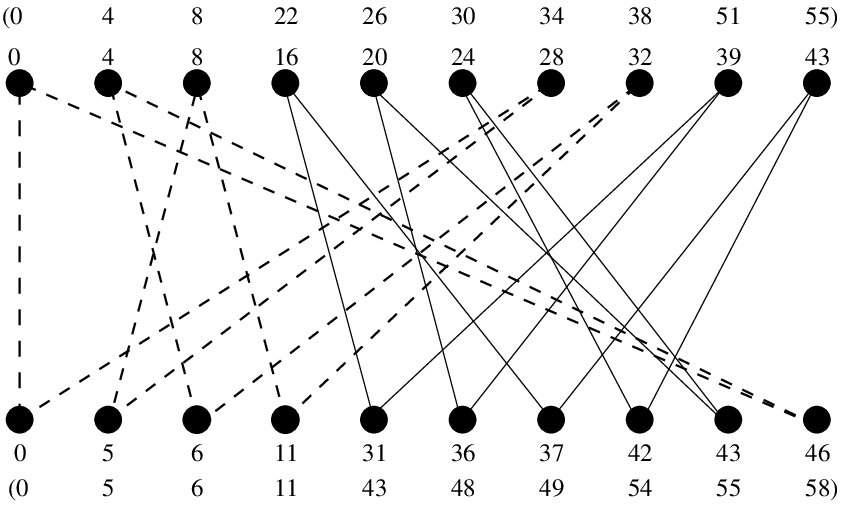}}
    \caption{Support matrix graph $G^{(0,1)}$ for the support matrix in (\ref{eq:supportmatrix1})  where the lower and upper layers correspond to the first and second rows, respectively. The cycle with dashed edges is the cycle $(v^{(0)}_{46},v^{(1)}_0,v^{(0)}_{0},v^{(1)}_{28},v^{(0)}_{5},v^{(1)}_{8},v^{(0)}_{11},v^{(1)}_{32},v^{(0)}_{6},v^{(1)}_{4},v^{(0)}_{46})$  (from (\ref{eq:cycle1})) of length $10$. The vertex labels in the parentheses correspond to the support matrix graph $G^{(0,1)}$  for the support matrix in (\ref{eq:supportmatrix2}) ($q=59$).}
    \label{fig:graph1}
\end{figure}

\begin{example} \label{ex:1}
Consider the case $q=47$ and $m=6$. Using a computer search, we have found a (minimal) codeword of weight $20$. The corresponding support matrix is
\begin{equation} \label{eq:supportmatrix1}
{\tiny {
\left[ \begin{smallmatrix}  
%
0&	42&	46&	5&	36&	46&	37&	31&	11&	5&	43&	6&	37&	0&	43&	42&	36&	31&	11&	6\\	
0&	43&	0&	8&	39&	4&	43&	39&	32&	28&	20&	32&	16&	28&	24&	24&	20&	16&	8&	4\\	
0&	44&	1&	11&	42&	9&	2&	0&	6&	4&	44&	11&	42&	9&	5&	6&	4&	1&	5&	2\\	
0&	45&	2&	14&	45&	14&	8&	8&	27&	27&	21&	37&	21&	37&	33&	35&	35&	33&	2&	0\\	
0&	46&	3&	17&	1&	19&	14&	16&	1&	3&	45&	16&	0&	18&	14&	17&	19&	18&	46&	45\\	
0&	0&	4&	20&	4&	24&	20&	24&	22&	26&	22&	42&	26&	46&	42&	46&	3&	3&	43&	43
%
%
%
\end{smallmatrix}  \right]}}
\end{equation}
and the support matrix graph $G^{(0,1)}$ (corresponding to the first two rows) is shown in Fig.~\ref{fig:graph1}. 
There is one distinct cycle in the graph containing the edge $(v^{(0)}_{46},v^{(1)}_0)$, namely the cycle 
\begin{equation} \label{eq:cycle1}
\left(v^{(0)}_{46},v^{(1)}_0,v^{(0)}_{0},v^{(1)}_{28},v^{(0)}_{5},v^{(1)}_{8},v^{(0)}_{11},v^{(1)}_{32},v^{(0)}_{6},v^{(1)}_{4},v^{(0)}_{46} \right)
\end{equation}
 (indicated with dashed edges in Fig.~\ref{fig:graph1}) of length $10$. 
Furthermore, for the support matrix
\begin{equation} \label{eq:supportmatrix2}
{\tiny {
\left[ \begin{smallmatrix}
%
0&	54&	58&	5&	48&	58&	49&	43&	11&	5&	55&	6&	49&	0&	55&	54&	48&	43&	11&	6\\	
0&	55&	0&	8&	51&	4&	55&	51&	38&	34&	26&	38&	22&	34&	30&	30&	26&	22&	8&	4\\	
0&	56&	1&	11&	54&	9&	2&	0&	6&	4&	56&	11&	54&	9&	5&	6&	4&	1&	5&	2\\	
0&	57&	2&	14&	57&	14&	8&	8&	33&	33&	27&	43&	27&	43&	39&	41&	41&	39&	2&	0\\	
0&	58&	3&	17&	1&	19&	14&	16&	1&	3&	57&	16&	0&	18&	14&	17&	19&	18&	58&	57\\	
0&	0&	4&	20&	4&	24&	20&	24&	28&	32&	28&	48&	32&	52&	48&	52&	56&	56&	55&	55	
%
%
\end{smallmatrix} \right]}}
\end{equation}
corresponding to a (minimal) codeword of weight $20$ for $q=59$ (and $m=6$), the corresponding cycle (also of length $10$) is 
\begin{equation} \label{eq:cycle2}
\left(v^{(0)}_{58},v^{(1)}_0,v^{(0)}_{0},v^{(1)}_{34},v^{(0)}_{5},v^{(1)}_{8},v^{(0)}_{11},v^{(1)}_{38},v^{(0)}_{6},v^{(1)}_{4},v^{(0)}_{58} \right).
\end{equation}
Thus, we get the same cycle lengths. The corresponding support matrix graph $G^{(0,1)}$ is shown in Fig.~\ref{fig:graph1} using the vertex labels in the parentheses. Continuing with the remaining pairs of rows, $(i,j)=(0,2),(0,3),\dots,(4,5)$, we get 
the same cycle lengths for cycles containing the edge $(v^{(i)}_{q-1+i \pmod{q}},v^{(j)}_{q-1+j \pmod{q}})$ or the edge $(v^{(i)}_{0},v^{(j)}_{0})$  for both support matrices. Thus, we would expect that there might exist a template support matrix whose instances (possibly column-permuted) for $q=47$ and $q=59$ are the support matrices in (\ref{eq:supportmatrix1}) and (\ref{eq:supportmatrix2}), respectively. 
%
%
\end{example}

\begin{algorithm}[tbp]
\begin{minipage}{\linewidth}
\caption{Template Support Matrix Inference}
\label{alg:tsmi}
\begin{algorithmic}[1]
\small
\STATE $/*$ Fill in entries in the candidate template support matrix in (\ref{eq:support_matrix_gen}) based on the support matrices of two (minimal) codewords $\mathbf{c}_1 \in \mathcal{C}(q_1,m)$ and $\mathbf{c}_2 \in \mathcal{C}(q_2,m)$, $q_2 > q_1$, of the same Hamming weight and with the same graphical cycle structure.\\
{\bf Input:} Row indices $i$ and $j$, a pair of cycles $(\mathbf{v}_1,\mathbf{v}_2)$ (of the same length $2l$) as defined in (\ref{eq:cycle1gen}) and (\ref{eq:cycle2gen}), and a positive integer $I$.\footnote{We will use $I=m-1$, although any value for $I$ can be used. However, using $I=m-1$ increases the likelihood of constructing a valid template support matrix.}\\
{\bf Output:} A (partial) candidate template support matrix as defined in (\ref{eq:support_matrix_gen}), and a (partial) permutation $\pi(\cdot)$. $*/$
\STATE Assign to $\mathcal{I}$ all integers in $\{1,\dots,I\}$. 
\FOR{$r \leftarrow 0$ to $2l-1$}
\STATE Find an index pair $(a,b)$ (which is also unique) such that $a \in \psi_1(v^{(\gamma)}_{\alpha_{1,r}}) \cap \psi_1(v^{(\delta)}_{\alpha_{1,r+1}})$ and $b \in \psi_2(v^{(\gamma)}_{\alpha_{2,r}}) \cap \psi_2(v^{(\delta)}_{\alpha_{2,r+1}})$, where $\gamma = i$ and $\delta=j$ if $r$ is even, and $\gamma = j$ and $\delta=i$ if $r$ is odd.
\STATE Solve the two systems of equations
\begin{align}
x^{(1)}_a+\gamma y^{(1)}_a \pmod {q_1} &= \alpha_{1,r} \notag \\
x^{(1)}_a+\delta y^{(1)}_a \pmod {q_1} &= \alpha_{1,r+1} \notag
\end{align}
and
\begin{align}
x^{(2)}_b+\gamma y^{(2)}_b \pmod {q_2} &= \alpha_{2,r} \notag \\
x^{(2)}_b+\delta y^{(2)}_b \pmod {q_2} &= \alpha_{2,r+1} \notag
\end{align}
\STATE Find the integers $k_x$ and $k_y$ in $\mathcal{I}$ that give the \emph{simplest} (defined below in the text) 
solutions (for $x$ and $y$, modulo $q_1 q_2$) to the two systems of congruences
\begin{align}
x &\equiv k_x \cdot x^{(1)}_a \pmod {q_1}  \notag \\
x &\equiv k_x \cdot x^{(2)}_b \pmod {q_2}  \notag 
\end{align}
and
\begin{align}
y &\equiv k_y \cdot y^{(1)}_a \pmod {q_1}  \notag \\
y &\equiv k_y \cdot y^{(2)}_b \pmod {q_2}  \notag 
\end{align}
%
%
\IF{$|x| \leq |x-q_1 q_2|$}
\STATE $\tilde{x} \leftarrow x \cdot k_x^{-1}$\\  
\ELSE 
\STATE $\tilde{x} \leftarrow (x-q_1 q_2) \cdot k_x^{-1}$\\  
\ENDIF
\IF{$|y| \leq |y-q_1 q_2|$}
\STATE $\tilde{y} \leftarrow y \cdot k_y^{-1}$ \\
\ELSE
\STATE $\tilde{y} \leftarrow (y -q_1 q_2) \cdot k_y^{-1}$ \\
\ENDIF
\IF{$x_a = \ast$ (and $y_a = \ast$)}
\STATE $x_a \leftarrow \tilde{x}$, $y_a \leftarrow \tilde{y}$, 
$\pi(b) \leftarrow a$, and go to Step 3.
%
%
\ELSIF{$x_a \neq \tilde{x}$ or $y_a \neq  \tilde{y}$}
\STATE an inconsistency has occurred. Exit.
\ENDIF
\ENDFOR
%
\end{algorithmic}
\end{minipage}
\end{algorithm} 

\subsection{Second Step: Inferring a Candidate Template Support Matrix} \label{sec:infer}

In this subsection, we consider the second step of the procedure, i.e., to infer a candidate template support matrix from two minimal codewords with the same graphical cycle structure. This is done by solving simple $2$-by-$2$ equation systems and congruences. 
We remark that this procedure will give a \emph{candidate} template support matrix, since we formally need to prove that the resulting matrix is a template support matrix. 

Now, let 
\begin{equation} \label{eq:cycle1gen}
\mathbf{v}_1 = 
(v_{\alpha_{1,0}}^{(i)}, v_{\alpha_{1,1}}^{(j)}, \dots, v_{\alpha_{1,2l-2}}^{(i)}, v_{\alpha_{1,2l-1}}^{(j)}, v_{\alpha_{1,2l}}^{(i)})
\end{equation}
 denote a cycle of length $2l$, where $\alpha_{1,0} = \alpha_{1,2l}$, in the support matrix graph $G_{\mathbf{c}_1}^{(i,j)}$ 
computed from a given minimal codeword $\mathbf{c}_1 \in \mathcal{C}(q_1,m)$. In a similar manner, we denote by 
\begin{equation} \label{eq:cycle2gen}
\mathbf{v}_2 = 
(v_{\alpha_{2,0}}^{(i)}, v_{\alpha_{2,1}}^{(j)}, \dots, v_{\alpha_{2,2l-2}}^{(i)}, v_{\alpha_{2,2l-1}}^{(j)}, v_{\alpha_{2,2l}}^{(i)})
\end{equation}
where $\alpha_{2,0} = \alpha_{2,2l}$,  a cycle of length $2l$ in the support matrix graph $G_{\mathbf{c}_2}^{(i,j)}$ 
computed from a given minimal codeword $\mathbf{c}_2 \in \mathcal{C}(q_2,m)$, where $q_2 > q_1$. We assume here that $\mathbf{c}_1$ and $\mathbf{c}_2$ have the same Hamming weight and also the same graphical cycle structure. Now, the purpose is to infer the entries in a matrix
\begin{equation} \label{eq:support_matrix_gen}
\left[ \begin{smallmatrix}
x_0 & x_1 & \cdots & x_{w-1} \\
x_0+y_0 & x_1+y_1 & \cdots &x_{w-1}+y_{w-1} \\
\cdots & \cdots & \cdots & \cdots \\
x_0+(m-1)y_0 & x_1+(m-1)y_1 & \cdots &x_{w-1}+(m-1)y_{w-1}
\end{smallmatrix}
\right]
\end{equation}
where $w$ is the Hamming weight of $\mathbf{c}_1$ and $\mathbf{c}_2$, 
such that the instances for $q=q_1$ and $q=q_2$ are the support matrices (possibly column-permuted)  of $\mathbf{c}_1$ and $\mathbf{c}_2$, respectively.

Algorithm~\ref{alg:tsmi} presents such an algorithm, where $\psi_1(v^{(i)}_{\alpha_1})$ (respectively $\psi_2(v^{(i)}_{\alpha_2})$) denotes the set of column indices of the support matrix of $\mathbf{c}_1$ (respectively $\mathbf{c}_2$) containing the entry ${\alpha_1}$ (respectively ${\alpha_2}$) in the $i$th row. 
%
All entries in the resulting matrix (after applying Algorithm~\ref{alg:tsmi}) should be reduced modulo $q$ to get an instance for a specific value of $q$. The algorithm works on two cycles of the same length, one from a support matrix graph of a minimal codeword $\mathbf{c}_1 \in \mathcal{C}(q_1,m)$ and the other from the corresponding support matrix graph of a minimal codeword $\mathbf{c}_2 \in \mathcal{C}(q_2,m)$, where $q_2 > q_1$. The purpose is to fill in the entries in a candidate template support matrix, which initially is filled  with erasures denoted by $\ast$. Furthermore, the algorithm also updates a permutation $\pi(\cdot)$ which gives the index mapping that should be applied to the columns of the support matrix of $\mathbf{c}_2$ to get the instance of the candidate template support matrix for $q=q_2$. The algorithm should run on pairs of cycles (both containing either the edge $(v^{(i)}_{q-1+i \pmod q},v^{(j)}_{q-1+j \pmod q})$ or the edge $(v^{(i)}_0,v^{(j)}_0)$ as the left-most or first edge in the cycle) until all entries are filled in. 

In Step 4 of the algorithm, an index pair $(a,b)$ is identified, where $a$ (respectively $b$) is the index of the column in the support matrix of $\mathbf{c}_1$ (respectively $\mathbf{c}_2$) containing $\alpha_{1,r}^{(\gamma)}$ (respectively $\alpha_{2,r}^{(\gamma)}$) as the $\gamma$th entry and $\alpha_{1,r+1}^{(\delta)}$ (respectively $\alpha_{2,r+1}^{(\delta)}$) as the $\delta$th entry. Later in Step 18 of the algorithm, these two indices are used to fill the permutation $\pi$ ($\pi(b) \leftarrow a$). Actually, the index pairs $(a,b)$ can be computed in a preprocessing stage before the algorithm has even been run, since they are available by simple cycle analysis.

In Steps 5 and 6 of the algorithm, we determine the entries in column $a$ (of the candidate template support matrix) based on the two cycles. In Step 5, we first determine the actual values for $x$ and $y$ modulo $q_1$ (denoted by $x_a^{(1)}$ and $y_a^{(1)}$, respectively) for column $a$ of the support matrix of $\mathbf{c}_1$, and then the corresponding values modulo $q_2$, now in column $b$ (denoted by $x_b^{(2)}$ and $y_b^{(2)}$, respectively), of the support matrix of $\mathbf{c}_2$.  Then, in Step 6, we find the \emph{simplest} solutions for $x$ and $y$ (modulo $q_1 q_2$), i.e., the solutions for $x$ and $y$, which also depend, respectively, on $k_x \in \mathcal{I}$ and $k_y \in \mathcal{I}$, that minimize, respectively, $\max(|k_x|, \min(|x|, |x-q_1 q_2|))$ and  $\max(|k_y|, \min(|y|, |y-q_1 q_2|))$. 
The solutions $x \cdot k_x^{-1}$ and $y \cdot k_y^{-1}$ both evaluate modulo $q$ (for $q=q_1$ and $q=q_2$) to the correct values as given by the support matrices of the codewords $\mathbf{c}_1$ and $\mathbf{c}_2$, respectively. 
%
Then, the entries for $x_a$ and $y_a$ are filled in the candidate template support matrix as defined in (\ref{eq:support_matrix_gen})
and 
as indicated in Steps 7 to 18. Note that in Steps 8 and 10 neither the inverse nor the product operation are performed and the formal string of three characters;  $x$ (or $x-q_1q_2$ in Step 10) (with a specific value inserted for $x$ (or $x-q_1 q_2$ evaluated for a specific value of $x$ in Step 10)), $\cdot$, and $k_x^{-1}$  (with a specific value inserted for $k_x$), is assigned to $\tilde{x}$. Of course, in the case of taking the inverse of $1$ or multiplying by $1$, the expression can be simplified by removing such terms. A similar comment applies to the assignments in Steps 13 and 15.  
Finally, we remark that using the simplest solutions,  as explained above, is to increase the likelihood that the candidate template support matrix is indeed a valid template matrix, and to find a candidate template support matrix with a nice/compact representation, which also makes it easier to prove analytically that all instances (possibly column-permuted) are indeed valid support matrices of codewords for all values of $q$ larger than or equal to some $\tilde{q}$, when $\tilde{q}$ is small (the third step of the heuristic). 
%
%
%
%
%
In any case, for any practical value of $\tilde{q}$, a simple and fast computer search can be used to prove whether or not the candidate template support matrix gives a valid support matrix for all values of $q_0 \leq q < \tilde{q}$, for some $q_0$. For details, see the proofs of Theorems~\ref{th:1} and \ref{th:2} in Sections~\ref{sec:bound} and \ref{sec:bound_m7}, respectively.  
%

Note that Algorithm~\ref{alg:tsmi} does in fact identify a one-to-one mapping (through the permutation $\pi(\cdot)$) between the columns of the support matrices of the codewords $\mathbf{c}_1 \in \mathcal{C}(q_1,m)$ and $\mathbf{c}_2 \in \mathcal{C}(q_2,m)$ by \emph{matching} cycles in the corresponding support matrix graphs. Then, the template values for $x$ and $y$ are established by \emph{matching} columns (and solving equations and congruences independently for each column) through this one-to-one mapping. It is in fact this particular one-to-one mapping (as opposed to an arbitrary mapping) that makes it possible for the resulting candidate template support matrix to have entries that appear an even number of times (the codeword case) or at least two times (the stopping set case) in each row.

In principle, one type of error condition can occur, i.e., we can exit in Step 20.  
This happens when a previous pair of cycles has determined the entries in column $a$ and then the current pair of cycles gives different values. 
If the algorithm exits in Step 20, we need to start from scratch by considering a  different pair of minimal codewords $\mathbf{c}_1 \in \mathcal{C}(q_1,m)$ and $\mathbf{c}_2 \in \mathcal{C}(q_2,m)$ of the same Hamming weight and with the same graphical cycle structure, or possibly the same pair if there are several possibilities for cycle pairs of the same length containing either the edge  $(v^{(i)}_{q-1+i \pmod q},v^{(j)}_{q-1+j \pmod q})$ or the edge $(v^{(i)}_0,v^{(j)}_0)$ for a given pair $(i,j)$,  and revert (back to erasures) all the entries filled in so far in the candidate template support matrix.\footnote{We remark that trying the same pair of codewords will be more important for the improved algorithm of Section~\ref{sec:improved} below.}   

In Step 5 of Algorithm~\ref{alg:tsmi}, two systems of equations need to be solved. They have the following solutions:
\begin{align}
x^{(1)}_a &= \alpha_{1,r}-\gamma (\delta-\gamma)^{-1} (\alpha_{1,r+1}-\alpha_{1,r}) \pmod {q_1} \notag \\
y^{(1)}_a &= (\delta-\gamma)^{-1}(\alpha_{1,r+1}-\alpha_{1,r}) \pmod {q_1} \notag \\
x^{(2)}_b &= \alpha_{2,r}-\gamma (\delta-\gamma)^{-1} (\alpha_{2,r+1}-\alpha_{2,r}) \pmod {q_2} \notag \\
y^{(2)}_b &= (\delta-\gamma)^{-1}(\alpha_{2,r+1}-\alpha_{2,r}) \pmod {q_2} \notag
\end{align}
which also gives the rationale behind the assignment to the set of integers $\mathcal{I}$ in Step 2 of the algorithm, 
 since the solutions involve a multiplication by $(\delta-\gamma)^{-1}$. 

\begin{figure*}[!t]
\normalsize \setcounter{mytempeqncnt}{\value{equation}}
\setcounter{equation}{12}
\begin{equation} \label{eq:supportm}
{\tiny {
\left[ \begin{smallmatrix}
%
%
%
%
%
%
0 & -5 & -1 & 5 & -11 & -1 & -10 & -16 & 11 & 5 & -4 & 6 & -10 & 0 & -4 & -5 & -11 & -16 & 11 & 6 \\
0 & -4 &  0 & 8 & -8 & 4 & -4 & -8 & 17 \cdot 2^{-1} & 9 \cdot 2^{-1} & -7 \cdot 2^{-1} & 17 \cdot 2^{-1} & -15 \cdot 2^{-1} & 9 \cdot 2^{-1} & 2^{-1} & 2^{-1} & -7 \cdot 2^{-1} & -15 \cdot 2^{-1} & 8 & 4\\
0 & -3 & 1 & 11 & -5 & 9 & 2 & 0 & 6 & 4 & -3 & 11 & -5 & 9 & 5 & 6 & 4 & 1 & 5 & 2 \\
0 & -2 & 2 & 14 & -2 & 14 & 8 & 8 & 7 \cdot 2^{-1} & 7 \cdot 2^{-1} & -5 \cdot 2^{-1} & 27 \cdot 2^{-1} & -5 \cdot 2^{-1} & 27 \cdot 2^{-1} & 19 \cdot 2^{-1} & 23 \cdot 2^{-1} & 23 \cdot 2^{-1} & 19 \cdot 2^{-1} & 2&  0\\
0 & -1 & 3 & 17 & 1 & 19 & 14 & 16 & 1 & 3 & -2 & 16 & 0 & 18 & 14 & 17 & 19 & 18 & -1 & -2 \\
0 & 0 & 4 & 20 & 4 & 24 & 20 & 24 & -3 \cdot 2^{-1} & 5 \cdot 2^{-1} & -3 \cdot 2^{-1} & 37 \cdot 2^{-1} & 5 \cdot 2^{-1} & 45 \cdot 2^{-1} & 37 \cdot 2^{-1} & 45 \cdot 2^{-1} & 53 \cdot 2^{-1} & 53 \cdot 2^{-1} & -4 & -4
\end{smallmatrix} \right]}}
\end{equation}
\setcounter{equation}{\value{mytempeqncnt}}
 \hrulefill
\vspace*{-2mm}
\end{figure*}

In Step 6 of Algorithm~\ref{alg:tsmi}, two systems of congruences need to be solved. They have the following solutions:
\begin{align}
x &= k_x (x_a^{(1)}+q_1 \cdot \kappa (x_b^{(2)}-x_a^{(1)})) \pmod {q_1 q_2} \label{eq:congu1} \\
y &= k_y (y_a^{(1)}+q_1 \cdot \kappa (y_b^{(2)}-y_a^{(1)})) \pmod {q_1 q_2} \label{eq:congu2} 
\end{align}
modulo $q_1 q_2$,
where $\kappa$ can be found using the extended Euclidean algorithm which yields integers $\kappa$ and $\eta$ such that $\kappa \cdot q_1 + \eta \cdot q_2 = \gcd(q_1,q_2)=1$.

Alternatively, in Steps~5 and 6 of Algorithm~\ref{alg:tsmi}, we can instead solve the two systems of congruences
\begin{align}
x+\gamma y &\equiv  \alpha_{1,r} \pmod {q_1}  \notag \\
x+\gamma y &\equiv  \alpha_{2,r} \pmod {q_2}  \notag 
\end{align}
and
\begin{align}
x+\delta y &\equiv \alpha_{1,r+1} \pmod {q_1}  \notag \\
x+\delta y &\equiv \alpha_{2,r+1} \pmod {q_2}  \notag 
\end{align}
for $x+\gamma y$ and $x+\delta y$, modulo $q_1 q_2$, 
and assign $(\gamma-\delta)^{-1} \cdot ((x+\delta y)\gamma- (x+\gamma y)\delta)$ and $(\gamma-\delta)^{-1} \cdot (x+\gamma y - (x+\delta y))$ to $x_a$ and $y_a$, respectively, in Step~18 of the algorithm. Here, both the inverse and the product operation are not performed, unless $(\gamma-\delta)$ is a divisor of  $(x+\delta y)\gamma- (x+\gamma y)\delta$ (for the assignment to $x_a$) or $x+\gamma y - (x+\delta y)$ (for the assignment to $y_a$). This will make the overall algorithm independent of the input parameter $I$, and will in fact be equivalent to running Algorithm~\ref{alg:tsmi} with $I=m-1$. We remark that using the simplest solutions from Step~6 is important for this equivalence. In the following, however, we will use the original version of Algorithm~\ref{alg:tsmi} with $I=m-1$. 

We will illustrate the procedure in Example~\ref{ex:2} below.

\begin{example} \label{ex:2}
Consider the two cycles in (\ref{eq:cycle1}) and (\ref{eq:cycle2}) for $q=47$ and $q=59$, respectively. Here, $i=0$ and $j=1$, and $\kappa = -5$ and $\eta=4$ (since $-5 \cdot 47 + 4 \cdot 59 = 1$). For $r=0$ (see Step 3 in Algorithm~\ref{alg:tsmi}), $\alpha_{1,r} = \alpha_{1,0} = 46$, $\alpha_{1,r+1} = \alpha_{1,1} = 0$, $\alpha_{2,r} = \alpha_{2,0} = 58$, $\alpha_{2,r+1} = \alpha_{2,1} = 0$, $\gamma=i=0$, and $\delta=j=1$. Since $46$ appears in the first row and $0$ in the second row of the third column (column index $2$) of the support matrix in (\ref{eq:supportmatrix1}), $a=2$. Similarly,  $b=2$, since $58$ appears in the first row and $0$ in the second row of the third column  of the support matrix in (\ref{eq:supportmatrix2}). This completes Step 4 of the algorithm, and we get the  solutions 
\begin{align}
x_{2}^{(1)} &= 46 - 0 \cdot (1-0)^{-1} (0-46) \pmod {47} = 46 \notag \\
y_{2}^{(1)} &= (1-0)^{-1} (0-46) \pmod {47} = 1 \notag \\
x_{2}^{(2)} &= 58 - 0 \cdot (1-0)^{-1} (0-58) \pmod {59} = 58 \notag \\
y_{2}^{(2)} &= (1-0)^{-1} (0-58) \pmod {59} = 1 \notag
\end{align}
in Step 5, 
from which we can calculate the following solutions for $x$ and $y$ in Step 6 (with $I=m-1=5$), using (\ref{eq:congu1}) and (\ref{eq:congu2}), respectively:
\begin{displaymath}
\begin{tabular}{r|rrrrrr}
$k_x$/$k_y$ & $1$ & $2$ & $3$ & $4$ & $5$ \\
\hline
$x$ & $2772$ & $2771$ & $2770$ & $2769$ & $2768$ \\
$x-q_1 q_2$ & $-1$ & $-2$&$-3$ &$-4$ & $-5$ \\
$y$ & $1$ & $2$&$3$ &$4$ & $5$ \\
$y-q_1 q_2$ & $-2772$ & $-2771$ & $-2770$ & $-2769$ & $-2768$
\end{tabular}.
\end{displaymath}
%
%
Thus, we can fill in $\pi(2)=2$, $x_{2}=-1$, and $y_{2}=1$ (corresponding to the values $k_x=1$ and $k_y=1$, which give the simplest solutions).

In a similar manner, for instance for $r=3$, we get $\alpha_{1,r} = \alpha_{1,3} = 28$, $\alpha_{1,r+1} = \alpha_{1,4} = 5$, $\alpha_{2,r} = \alpha_{2,3} = 34$, $\alpha_{2,r+1} = \alpha_{2,4} = 5$, $\gamma=j=1$, and $\delta=i=0$. For this case we have $a=9$ and $b=9$ (from Step 4), and the solutions
\begin{align}
x_{9}^{(1)} &= 28 - 1 \cdot (0-1)^{-1} (5-28) \pmod {47} = 5 \notag \\
y_{9}^{(1)} &= (0-1)^{-1} (5-28) \pmod {47} = 23 \notag \\
x_{9}^{(2)} &= 34 - 1 \cdot (0-1)^{-1} (5-34) \pmod {59} = 5 \notag \\
y_{9}^{(2)} &= (0-1)^{-1} (5-34) \pmod {59} = 29 \notag
\end{align}
 in Step 5, 
from which we can calculate the following solutions for $x$ and $y$ in Step 6 (with $I=m-1=5$), using (\ref{eq:congu1}) and (\ref{eq:congu2}), respectively:
\begin{displaymath}
\begin{tabular}{r|rrrrrr}
$k_x$/$k_y$ & $1$ & $2$ & $3$ & $4$ & $5$ \\
\hline
$x$ & $5$ & $10$ & $15$ & $20$ & $25$ \\
$x-q_1 q_2$ & $-2768$ & $-2763$ & $-2758$ & $-2753$ & $-2748$ \\
$y$ & $1386$ & $2772$ & $1385$ & $2771$ & $1384$ \\
$y-q_1 q_2$ & $-1387$ & $-1$ & $-1388$ & $-2$ & $-1389$ 
\end{tabular}.
\end{displaymath}
%
Thus, we can fill in $\pi(9)=9$, $x_{9}=5$, and $y_{9}=-2^{-1}$ (corresponding to the values $k_x=1$ and $k_y=2$, which give the simplest solutions). 

Continuing with the rest of the values for $r$ (see Step 3 in Algorithm~\ref{alg:tsmi}) a total of $10$ (the cycle length) columns of the candidate template support matrix can be determined. To determine the rest of the entries in the  matrix, other cycle pairs must be considered. 
%
%
For instance, by looking at the graphs $G^{(0,2)}$, we find the cycles
\begin{equation} \notag 
\left(v^{(0)}_{46},v^{(2)}_1,v^{(0)}_{31},v^{(2)}_{0},v^{(0)}_{0},v^{(2)}_{9},v^{(0)}_{46} \right)
\end{equation}
and
\begin{equation} \notag 
\left(v^{(0)}_{58},v^{(2)}_1,v^{(0)}_{43},v^{(2)}_{0},v^{(0)}_{0},v^{(2)}_{9},v^{(0)}_{58} \right)
\end{equation}
for $q=47$ and $q=59$, respectively. Choose $r=1$, from which we get  $\alpha_{1,r} = \alpha_{1,1} = 1$, $\alpha_{1,r+1} = \alpha_{1,2} = 31$, $\alpha_{2,r} = \alpha_{2,1} = 1$, $\alpha_{2,r+1} = \alpha_{2,2} = 43$, $\gamma=j=2$, and $\delta=i=0$. For this case we have $a=17$ and $b=17$ (from Step 4), and the solutions
\begin{align}
x_{17}^{(1)} &= 1 - 2 \cdot (0-2)^{-1} (31-1) \pmod {47} = 31 \notag \\
y_{17}^{(1)} &= (0-2)^{-1} (31-1) \pmod {47} = 32 \notag \\
x_{17}^{(2)} &= 1 - 2 \cdot (0-2)^{-1} (43-1) \pmod {59} = 43 \notag \\
y_{17}^{(2)} &= (0-2)^{-1} (43-1) \pmod {59} = 38 \notag
\end{align}
in Step 5, 
from which we can calculate the following solutions for $x$ and $y$ in Step 6 (with $I=m-1=5$), using (\ref{eq:congu1}) and (\ref{eq:congu2}), respectively:
\begin{displaymath}
\begin{tabular}{r|rrrrrr}
$k_x$/$k_y$ & $1$ & $2$ & $3$ & $4$ & $5$ \\
\hline
$x$ & $2757$ & $2741$ & $2725$ & $2709$ & $2693$ \\
$x-q_1 q_2$ & $-16$ & $-32$ & $-48$ & $-64$ & $-80$ \\
$y$ & $1395$ & $17$ & $1412$ & $34$ & $1429$ \\
$y-q_1 q_2$ & $-1378$ & $-2756$ & $-1361$ & $-2739$ & $-1344$
\end{tabular}.
\end{displaymath}
Thus, we can fill in $\pi(17)=17$, $x_{17}=-16$, and $y_{17}=17 \cdot 2^{-1}$ (corresponding to the values $k_x=1$ and $k_y=2$, which give the simplest solutions). 
 Continuing (by considering more cycle pairs) we can determine the rest of the columns,  and we end up with the candidate template support matrix shown in (\ref{eq:supportm}) at the top of the page, where all entries should be reduced modulo $q$ to get an instance for a specific value of $q$. The remaining detailed calculations are omitted for brevity.

\end{example}


\subsection{Third Step: A Formal Proof} \label{sec:formal}

The third step is showing that the candidate template support matrix is indeed a valid template support matrix for some parameter $q_0$, i.e., the instances for $q \geq q_0$ (possibly column-permuted) are all valid support matrices of codewords from $\mathcal{C}(q,m)$. In fact it is sufficient (to prove an upper bound on the minimum distance) to show that the instances for $q \geq q_0$ (possibly column-permuted) all contain as submatrices valid support matrices of codewords from $\mathcal{C}(q,m)$. In the case an instance (possibly column-permuted) contains as a proper submatrix a valid support matrix of a codeword, the established upper bound is obviously not tight. In particular, we need to show, for any value of $q \geq q_0$, for some $q_0$, that
\begin{enumerate}
\item all entries in a row occur an even number of times,  
\item all columns in the matrix are in fact valid columns in an array LDPC code parity-check matrix, and 
\item the \emph{column-reduced} matrix modulo $q$, which is obtained by removing all pairs of identical columns, is nonempty. 
For instance, if a column vector appears an odd number of times in the candidate support matrix, then all but one of these columns are removed for the column-reduced version, and if a column vector appears an even number of times, then all of these columns are removed for the column-reduced version. Note that the column-reduced matrix (when conditions 1) and 2) above are satisfied for the non-column-reduced version) is always a valid (possibly column-permuted or even empty) support matrix, since the removal of a pair of identical columns does not violate the first condition (and obviously not the second condition) above. This third condition is satisfied, for instance, if at least two columns are distinct modulo $q$ and appear an odd number of times. 
%
%
\end{enumerate}
Note that the second condition above will always be satisfied if Algorithm~\ref{alg:tsmi} indeed produces a complete candidate template support matrix, 
since by construction all columns are of the form in (\ref{eq:form}), for some $x$ and $y$, and all possible values for $x$ and $y$ will give a valid column 
(see the discussion following (\ref{eq:form})). Thus, only the first and third conditions above need to be explicitly verified if in fact the candidate template support matrix was produced by Algorithm~\ref{alg:tsmi}. 

Finally, we remark that complete formal proofs for the three conditions in the list above will be provided below in Section~\ref{sec:bound} for the case where $m=6$ and in Section~\ref{sec:bound_m7} for the case where $m=7$.


\subsection{Adaption to the Stopping Set Case} \label{sec:stopping}

In this subsection, we briefly describe how the approach changes when it is used for deriving an upper bound on the stopping distance.

The first step of the approach does not change at all, since it is based on the concepts of support matrices and support matrix graphs. Instead of considering the support matrix of a codeword, we consider the support matrix of a stopping set. Also, the operation of Algorithm~\ref{alg:tsmi} is the same. Instead of filling in entries in the candidate template support matrix in (\ref{eq:support_matrix_gen}) based on the support matrices of two (minimal) codewords  of the same Hamming weight and with the same graphical cycle structure, we  fill in entries in (\ref{eq:support_matrix_gen}) based on the support matrices of two (minimal) stopping sets  of the same size and with the same graphical cycle structure.

For the third step, condition 2) in Section~\ref{sec:formal} is always satisfied for the same reason as in the codeword case. Thus, only the first and third conditions (of Section~\ref{sec:formal}) need to be explicitly verified (as in the codeword case) if in fact the candidate template support matrix was produced by Algorithm~\ref{alg:tsmi}.  Note that the first condition should be modified to fit the stopping set case. Instead of requiring that all entries in a row occur an even number of times, all entries should appear at least two times in each row.  As for the first condition, the third condition should also be modified to fit the stopping set case. Instead of requiring, for instance,  that at least two columns are distinct modulo $q$ and appear an odd number of times, we can run the following column-removal algorithm on the candidate support matrix. Let $\tilde{\mathbf{H}}_q$ denote the candidate template support matrix modulo $q$ for some fixed value of $q$. $^\dagger$If there are no repeated columns in $\tilde{\mathbf{H}}_q$, then exit. Otherwise, locate a column vector that appears a multiple number of times in $\tilde{\mathbf{H}}_q$ and remove all but one of these columns from $\tilde{\mathbf{H}}_q$. If the first condition is violated for the resulting matrix  $\tilde{\mathbf{H}}_q$, then remove also the remaining column (of the located repeated columns) from $\tilde{\mathbf{H}}_q$. Repeat from $^\dagger$ if the resulting matrix  $\tilde{\mathbf{H}}_q$ satisfies the first condition. Otherwise, terminate the algorithm. 
%
Now, the third condition is satisfied (by definition) if and only if the resulting matrix $\tilde{\mathbf{H}}_q$ (after running the algorithm above) is nonempty and satisfies the first condition. We remark that a different processing order on the set of repeated columns may produce a different matrix $\tilde{\mathbf{H}}_q$ at the end of the algorithm. Thus, in case the resulting matrix $\tilde{\mathbf{H}}_q$ is nonempty and does not satisfy the third (or, equivalently, the first) condition outlined above, the algorithm can be run again using a different processing order on the set of repeated columns, ultimately trying all possible processing orders. Note that in the special case of no repeated columns in the original candidate template support matrix modulo $q$ for any fixed $q$, the algorithm will remove no columns and the third condition will automatically be satisfied due to the first condition. Also, note that running the above algorithm in the codeword case (using any processing order on the set of repeated columns) will produce the column-reduced candidate support matrix (as defined above in Section~\ref{sec:formal}), and the first condition will always be satisfied for the resulting matrix. Thus, in the codeword case, we get the condition that the column-reduced matrix should be nonempty.

Finally, we remark that an efficient algorithm to find small-size stopping sets is required.

\subsection{Applicability}

The heuristic outlined above in Sections~\ref{sec:structure} through \ref{sec:formal} is very general and can be applied for any pair of values $(q,m)$. However, the difficult part is finding low-weight/small-size candidate codewords/stopping sets for different values of $q$, which is increasingly difficult when $m$ grows, since the minimum/stopping distance increases with $m$. For this we have used the algorithm in \cite{ros09,ros12}, and the minimum distance probabilistic algorithm in \cite{tom07}.

In this work, we have applied the heuristic for $m=6$ and $m=7$, but remark that it will easily provide the upper bounds $d(q,4) \leq 10$ and $d(q,5) \leq 12$ which can be found in the literature \cite{sug08}. In fact, the proposed approach resembles the approach of Sugiyama and Kaji in \cite{sug08}. Also, in \cite{sug08}, support matrices of actual codewords for different values of $q$ ($m$ fixed to either $4$ or $5$) are used to identify what is called ``\emph{cancel-out patterns}'' in \cite{sug08} (each distinct entry in a row in a support matrix occurs an even number of times). However, they do not connect the support matrices to graphs and cycles in graphs in a systematic way as we do here. As we will show below in Section~\ref{sec:bound_m7}, we can also deal with pairs of codewords which do not share the same ``cancel-out patterns'' (as opposed to the basic approach from \cite{sug08}). This is important when $m$ grows. Hence, we are able to deal with larger values of $m$.

\subsection{Improved Algorithm} \label{sec:improved}

The basic algorithm from Sections~\ref{sec:structure} and \ref{sec:infer} can be improved in the sense of increasing its probability of success, i.e., of finding a valid template support matrix. The key observation in this respect is that even though the two codewords $\mathbf{c}_1 \in \mathcal{C}(q_1,m)$ and $\mathbf{c}_2 \in \mathcal{C}(q_2,m)$ do not have the same graphical cycle structure, their support matrices (possibly column-permuted) may still be instances of the same template matrix. The reason is that different entries in the template matrix may reduce to the same value modulo $q$ for different values of $q$. This typically happens when either $q_1$ or $q_2$ is small. A simple way to deal with such scenarios is by relaxing the condition that $\mathbf{c}_1$ and $\mathbf{c}_2$ should have \emph{exactly} the same graphical cycle structure. In particular, it may be sufficient to require that the \emph{minimum} cycle length of all cycles  containing the edge $(v^{(i)}_{q-1+i \pmod q},v^{(j)}_{q-1+j \pmod q})$ and the \emph{minimum} cycle length of all cycles containing the edge $(v^{(i)}_0,v^{(j)}_0)$ are the same for both support matrix graphs $G_{\mathbf{c}_1}^{(i,j)}$ and $G_{\mathbf{c}_2}^{(i,j)}$, $0 \leq i < j < m$, and then run Algorithm~\ref{alg:tsmi} on such pairs of cycles (which have the same length).


\begin{figure*}[t]
\normalsize \setcounter{mytempeqncnt}{\value{equation}}
\setcounter{equation}{16}
\begin{equation} \label{eq:supportm7}
{\tiny {
\begin{split}
&\left[ \begin{smallmatrix}
0 & 3 \cdot 2^{-1} & 0 & -9 \cdot 2^{-1} & -7 \cdot 2^{-1} & -1 & -11 \cdot 2^{-1} & -5 & 2 & -2 & -5 & -1 & 2 &  5 \cdot 2^{-1} & 2^{-1} &  3 \cdot 2^{-1} & -3 & -2 & -9 \cdot 2^{-1} & \\
0 & 3 \cdot 2^{-1} & 1 & -7 \cdot 2^{-1} & -5 \cdot 2^{-1} & 0 & -7 \cdot 2^{-1} & -3 & 9\cdot 2^{-2} & -7 \cdot 2^{-2} & -15\cdot 2^{-2} & 2^{-2} & 3 \cdot 2^{-1} & 2 &  1 &  2 &   -5 \cdot 2^{-1} & -3 \cdot 2^{-1} & -3  & \\
0 & 3 \cdot 2^{-1} & 2 & -5 \cdot 2^{-1} & -3 \cdot 2^{-1} &  1 & -3 \cdot 2^{-1} &  -1 & 5 \cdot 2^{-1} &  -3 \cdot 2^{-1} & -5 \cdot 2^{-1} & 3 \cdot 2^{-1} &  1 &   3 \cdot 2^{-1} & 3 \cdot 2^{-1} & 5 \cdot 2^{-1} & -2 &  -1 &   -3 \cdot 2^{-1} & \\
0 & 3 \cdot 2^{-1} & 3 & -3 \cdot 2^{-1} & -2^{-1} &   2 &  2^{-1}  &   1 & 11\cdot 2^{-2} &  -5\cdot 2^{-2} & -5\cdot 2^{-2} &  11\cdot 2^{-2} & 2^{-1} &  1 & 2 &   3 &   -3 \cdot 2^{-1} & -2^{-1} &  0  & \\ 
0 & 3 \cdot 2^{-1} & 4 & -2^{-1} &   2^{-1}  &  3 &  5 \cdot 2^{-1} &   3 & 3 &    -1 &    0 &    4 &   0 &   2^{-1} &  5 \cdot 2^{-1} & 7 \cdot 2^{-1} & -1 &   0 &    3 \cdot 2^{-1}  & \\
0 & 3 \cdot 2^{-1} & 5 &  2^{-1} &   3 \cdot 2^{-1} &  4 &  9 \cdot 2^{-1} &   5 & 13\cdot 2^{-2} & -3\cdot 2^{-2} &  5\cdot 2^{-2} &  21 \cdot 2^{-2} & -2^{-1} & 0 &   3 &   4 &  -2^{-1}  & 2^{-1}  &  3 & \\
0 & 3 \cdot 2^{-1} & 6 &  3 \cdot 2^{-1} &  5 \cdot 2^{-1} &  5 &  13 \cdot 2^{-1} & 7 & 7 \cdot 2^{-1} &  -2^{-1} &   5 \cdot 2^{-1} &  13 \cdot 2^{-1} & -1 & -2^{-1} &  7 \cdot 2^{-1} & 9 \cdot 2^{-1} & 0 &    1 &    9 \cdot 2^{-1}  &
\end{smallmatrix} \right.  \\ %
& \left. \begin{smallmatrix}
& -3   & 2^{-1} &    5 \cdot 2^{-1} & -11 \cdot 2^{-1} & -7 \cdot 2^{-1}  \\
& -3 \cdot 2^{-1} & 2^{-2} &   9\cdot 2^{-2} & -15\cdot 2^{-2} & -7\cdot 2^{-2} \\
& 0 &    0 &    2 &   -2 &    0 \\
\hspace{10.0cm} &    3 \cdot 2^{-1} &  -2^{-2} &   7\cdot 2^{-2} & -2^{-2} &   7\cdot 2^{-2} \\
&  3 &    -2^{-1} &  3 \cdot 2^{-1} & 3 \cdot 2^{-1} &   7 \cdot 2^{-1} \\ 
&     9 \cdot 2^{-1} &  -3\cdot 2^{-2} & 5 \cdot 2^{-2} & 13 \cdot 2^{-2} &  21 \cdot 2^{-2}  \\
&  6 &   -1 &   1 &   5 &    7  \end{smallmatrix} \right] 
\end{split} }}
\end{equation}
\setcounter{equation}{\value{mytempeqncnt}}
 \hrulefill
\vspace*{-2mm}
\end{figure*}

\section{Upper Bound on $d(q,6)$} \label{sec:bound}

By using the heuristic from Section~\ref{sec:algo}, 
we have found the \emph{candidate} template support matrix in (\ref{eq:supportm}), in which all entries should be reduced modulo $q$. At this stage we emphasize that this is a \emph{candidate} template support matrix, since we need to formally prove that the matrix is a template support matrix.  In particular, we have used the procedure from Section~\ref{sec:infer} to infer the  matrix in (\ref{eq:supportm})  from the codewords of Example~\ref{ex:1}, which have the same graphical cycle structure. Also, in Example~\ref{ex:2}, some of the columns in the matrix in (\ref{eq:supportm}) were explicitly determined. The rest of the columns can be determined in a similar manner. Details are omitted for brevity. 

We can now prove the following theorem.

\begin{theorem} \label{th:1}
The minimum distance $d(q,6)$ is upper-bounded by $20$ for $q > 11$.
\end{theorem}

\begin{IEEEproof}
The proof is based on the candidate template support matrix in (\ref{eq:supportm}). As explained in Section~\ref{sec:formal}, there are three conditions that need to be verified. Also, if the candidate template support matrix was indeed produced by Algorithm~\ref{alg:tsmi}, only the first and third conditions need to be explicitly verified. For completeness and for providing a formal proof, we will however verify all three conditions. Obviously, computing an upper bound on the minimum distance from a template support matrix based on a codeword is easy; the upper bound is just the number  of columns in the matrix. Thus, establishing that the matrix in  (\ref{eq:supportm}) is a valid template support matrix, in the sense that all instances (possibly column-permuted) for $q > 11$ contain the support matrix of a codeword as a submatrix, establishes the upper bound of $20$, since there are $20$ columns in the matrix.

It is easy to verify  that each entry in each row of the matrix appears exactly twice, which means that the result is true if for any value of $q > 11$ 
\begin{enumerate}
\item[2)] all columns in the matrix are in fact valid columns in an array LDPC code parity-check matrix, and
\item[3)] at least two columns are distinct modulo $q$ and appear an odd number of times. 
\end{enumerate}

Since all columns in the matrix in (\ref{eq:supportm})  are of the form in (\ref{eq:form}), it follows that they are all valid columns in an array LDPC code parity-check matrix (see the discussion following (\ref{eq:form})). 
%
In particular, the values for $x,y$ for the first $6$ columns are

\begin{displaymath}
\begin{tabular}{c|cccccc}
$x$&$0$&$-5$&$-1$&$5$&$-11$&$-1$\\
\hline
$y$&$0$&$1$&$1$&$3$&$3$&$5$
\end{tabular}.
\end{displaymath}

For the third part of the proof, we need to show, for any value of $q > 11$, that there exist (at least two) columns in the candidate template support matrix  which are not identical modulo $q$ and appear an odd number of times. This is simple  (and very fast) to verify by a computer search for any finite value of $q$ that would be of any practical value.  It is only for large values of $q$ that the theoretical proof below is needed.

Note that the maximum absolute value of the entries in the first row of the  matrix in (\ref{eq:supportm})  is $16$. 
Thus, the only possibility for repeated columns,  when $q > 2 \cdot 16 = 32$, is for two \emph{neighboring} columns (with identical entries in the first row) to be the same. 
However, by looking at the third row in the matrix, this possibility can be ruled out by requiring that $q$ is larger than twice the maximum absolute value of the entries in the third row, i.e., by requiring $q > 2 \cdot 11 = 22$. 
In summary, it follows that there are no identical columns in the matrix in (\ref{eq:supportm})  if $q > \max(32,22)=32$. 
Furthermore, for values of $11 < q < 32$, it can be verified numerically that there are no repeated columns in (\ref{eq:supportm}), and the result follows. 
%
\end{IEEEproof}

\setcounter{equation}{13}
We remark that for $q=7$, the matrix in (\ref{eq:supportm}) reduces to
\begin{equation} \label{eq:supportmq7}
\left[ \begin{smallmatrix}
 0&  2&  6&  5&  3&  6&  4&  5&  4&  5&  3&  6&  4&  0&  3&  2&  3&  5&  4&  6\\ 
 0&  3&  0&  1&  6&  4&  3&  6&  5&  1&  0&  5&  3&  1&  4&  4&  0&  3&  1&  4 \\
 0&  4&  1&  4&  2&  2&  2&  0&  6&  4&  4&  4&  2&  2&  5&  6&  4&  1&  5&  2 \\
 0&  5&  2&  0&  5&  0&  1&  1&  0&  0&  1&  3&  1&  3&  6&  1&  1&  6&  2&  0 \\
 0&  6&  3&  3&  1&  5&  0&  2&  1&  3&  5&  2&  0&  4&  0&  3&  5&  4&  6&  5 \\
 0&  0&  4&  6&  4&  3&  6&  3&  2&  6&  2&  1&  6&  5&  1&  5&  2&  2&  3&  3 
\end{smallmatrix} \right].
\end{equation}
We observe that there are indeed some identical columns when $q=7$. 
However, the bound in Theorem~\ref{th:1} is still valid, since these columns can just be removed from (\ref{eq:supportmq7}) and we will end up in the valid (but column-permuted) support matrix
\begin{equation}  \label{eq:supportmq7red}
\left[ \begin{smallmatrix}
%
 0&  2&  6&    3&       5&  4&        6&     0&  3&  2&    5&  4  \\ 
 0&  3&  0&    6&        6&  5&        5&     1&  4&  4&    3&  1   \\
 0&  4&  1&    2&        0&  6&        4&     2&  5&  6&    1&  5   \\
 0&  5&  2&    5&        1&  0&        3&     3&  6&  1&    6&  2   \\
 0&  6&  3&    1&        2&  1&        2&     4&  0&  3&    4&  6   \\
 0&  0&  4&    4&        3&  2&        1&     5&  1&  5&    2&  3   
\end{smallmatrix} \right]
\end{equation}
which corresponds to a codeword of weight $12$, but the bound $d(7,6) \leq 20$ is of course not tight in this case. In fact, we found by exhaustive search that the codeword corresponding to the matrix in (\ref{eq:supportmq7red}) is indeed a minimum-weight codeword. Similarly, for $q=11$, the matrix in (\ref{eq:supportm}) reduces to
\begin{equation} \label{eq:supportmq11red}
\left[ \begin{smallmatrix}
 0&     6&    10&     5&     10&    1&    5&     7&     1&     0&     7&     6&     0&     6&     0&     6\\
 0&     7&     0&     8&     4&     7&   10&     2&     9&    10&     6&     6&     2&     9&     8&     4\\
 0&     8&     1&     0&     9&     2&    4&     8&     6&     9&     5&     6&     4&     1&     5&     2\\
 0&     9&     2&     3&     3&     8&    9&     3&     3&     8&     4&     6&     6&     4&     2&     0\\
 0&    10&     3&     6&     8&     3&    3&     9&     0&     7&     3&     6&     8&     7&    10&     9\\
 0&     0&     4&     9&     2&     9&    8&     4&     8&     6&     2&     6&    10&    10&     7&     7
\end{smallmatrix} \right]
\end{equation}
after removing pairs of identical columns, which corresponds to a codeword of weight $16$. As for $q=7$, the bound in Theorem~\ref{th:1} is still valid, but not tight in this case as well. By running an exhaustive search, we found that the codeword corresponding to the matrix in (\ref{eq:supportmq11red}) is in fact a minimum-weight codeword.


Finally, we remark that the template support matrix in (\ref{eq:supportm}) for $q=7$, $11$, $13$, $17$, and $19$ does not give instances with columns in the order as implied by the parity-check matrix in (\ref{eq:pcmatrix}). This can easily be seen from the sequence of $y$-values for the matrix in (\ref{eq:supportm}), which should be nondecreasing. 
Furthermore, if two $y$-values are the same, then the corresponding sequence of $x$-values  should be nondecreasing. For $q > 19$, it can easily be proved that the order is always according to  (\ref{eq:pcmatrix}).  However, as argued previously, this is not important (independent of the order, a support matrix will represent the same codeword/stopping set).


\section{Upper Bound on $d(q,7)$} \label{sec:bound_m7}

For the case $m=7$ we have found, using the algorithm from \cite{tom07}, the support matrices
\begin{equation}  \notag 
\left[ \begin{smallmatrix}
0 & 13 & 0 & 7 & 8 & 22 & 6 & 18 & 2 & 21 & 18 & 22 & 2 & 14 & 12 & 13 & 20 & 21 & 7 & 20 & 12 & 14 & 6 & 8 \\
0 & 13 & 1 & 8 & 9 & 0 & 8 & 20 & 8 & 4 & 2 & 6 & 13 & 2 & 1 & 2 & 9 & 10 & 20 & 10 & 6 & 8 & 2 & 4 \\
0 & 13 & 2 & 9 & 10 & 1 & 10 & 22 & 14 & 10 & 9 & 13 & 1 & 13 & 13 & 14 & 21 & 22 & 10 & 0 & 0 & 2 & 21 & 0 \\
0 & 13 & 3 & 10 & 11 & 2 & 12 & 1 & 20 & 16 & 16 & 20 & 12 & 1 & 2 & 3 & 10 & 11 & 0 & 13 & 17 & 19 & 17 & 19 \\
0 & 13 & 4 & 11 & 12 & 3 & 14 & 3 & 3 & 22 & 0 & 4 & 0 & 12 & 14 & 15 & 22 & 0 & 13 & 3 & 11 & 13 & 13 & 15 \\
0 & 13 & 5 & 12 & 13 & 4 & 16 & 5 & 9 & 5 & 7 & 11 & 11 & 0 & 3 & 4 & 11 & 12 & 3 & 16 & 5 & 7 & 9 & 11 \\
0 & 13 & 6 & 13 & 14 & 5 & 18 & 7 & 15 & 11 & 14 & 18 & 22 & 11 & 15 & 16 & 0 & 1 & 16 & 6 & 22 & 1 & 5 & 7
\end{smallmatrix} \right]
\end{equation}
and
\begin{equation}  \notag 
\left[ \begin{smallmatrix}
0 & 16 & 0 & 10 & 11 & 28 & 9 & 24 & 15 & 17 & 9 & 11 & 2 & 17 & 15 & 16 & 26 & 27 & 10 & 26 & 2 & 27 & 24 & 28 \\
0 & 16 & 1 & 11 & 12 & 0 & 11 & 26 & 22 & 24 & 18 & 20 & 16 & 2 & 1 & 2 & 12 & 13 & 26 & 13 & 24 & 20 & 18 & 22 \\
0 & 16 & 2 & 12 & 13 & 1 & 13 & 28 & 0 & 2 & 27 & 0 & 1 & 16 & 16 & 17 & 27 & 28 & 13 & 0 & 17 & 13 & 12 & 16 \\
0 & 16 & 3 & 13 & 14 & 2 & 15 & 1 & 7 & 9 & 7 & 9 & 15 & 1 & 2 & 3 & 13 & 14 & 0 & 16 & 10 & 6 & 6 & 10 \\
0 & 16 & 4 & 14 & 15 & 3 & 17 & 3 & 14 & 16 & 16 & 18 & 0 & 15 & 17 & 18 & 28 & 0 & 16 & 3 & 3 & 28 & 0 & 4 \\
0 & 16 & 5 & 15 & 16 & 4 & 19 & 5 & 21 & 23 & 25 & 27 & 14 & 0 & 3 & 4 & 14 & 15 & 3 & 19 & 25 & 21 & 23 & 27 \\
0 & 16 & 6 & 16 & 17 & 5 & 21 & 7 & 28 & 1 & 5 & 7 & 28 & 14 & 18 & 19 & 0 & 1 & 19 & 6 & 18 & 14 & 17 & 21 
\end{smallmatrix} \right]
\end{equation}
 of (minimal) codewords $\mathbf{c}_1$ and  $\mathbf{c}_2$ of weight $24$ for $q=23$ and $q=29$, respectively. 
For instance, note that in the matrix for $q=23$ (the first matrix) the entries $5$ and $11$ appear four times in the second-to-last row, while in the matrix for $q=29$ (the second matrix) all entries appear twice in the second-to-last row. In the last row, however, all entries appear twice for both matrices. 
As a consequence, there are two different cycles 
\begin{displaymath}
\left(v^{(5)}_{0},v^{(6)}_0,v^{(5)}_{11},v^{(6)}_{7},v^{(5)}_{5},v^{(6)}_{11},v^{(5)}_{0} \right)
\end{displaymath}
and
\begin{displaymath}
\left(v^{(5)}_{0},v^{(6)}_0,v^{(5)}_{11},v^{(6)}_{22},v^{(5)}_{5},v^{(6)}_{11},v^{(5)}_{0} \right)
\end{displaymath}
of length $6$ and one cycle 
\begin{displaymath}
\left(v^{(5)}_{0},v^{(6)}_0,v^{(5)}_{11},v^{(6)}_{18},v^{(5)}_{16},v^{(6)}_{6},v^{(5)}_{5},v^{(6)}_{11},v^{(5)}_{0} \right)
\end{displaymath}
of length $8$  containing the edge $(v^{(5)}_{0},v^{(6)}_{0})$  in the support matrix graph $G_{\mathbf{c}_1}^{(5,6)}$ (corresponding to the first matrix), while there is only a single such cycle
\begin{displaymath}
\left(v^{(5)}_{0},v^{(6)}_0,v^{(5)}_{14},v^{(6)}_{28},v^{(5)}_{21},v^{(6)}_{14},v^{(5)}_{0} \right)
\end{displaymath}
 (of length $6$) in the support matrix graph  $G_{\mathbf{c}_2}^{(5,6)}$ (corresponding to the second matrix). Hence, the codewords $\mathbf{c}_1$ and $\mathbf{c}_2$ do \emph{not} have the same graphical cycle structure, and they also  have different ``cancel-out patterns''. 
%
Note, however, that the \emph{minimum} cycle lengths are the same, and this is also the case for all the other pairs of  graphs  $G_{\mathbf{c}_1}^{(i,j)}$ and  $G_{\mathbf{c}_2}^{(i,j)}$, $0 \leq i < j < m$, although for several values of $(i,j)$ the graph $G_{\mathbf{c}_1}^{(i,j)}$ contains more cycles of longer lengths than the graph $G_{\mathbf{c}_2}^{(i,j)}$.  Following the discussion in Section~\ref{sec:improved}, we may apply Algorithm~\ref{alg:tsmi}, which infers the candidate template support matrix shown in (\ref{eq:supportm7}) at the top of the page. Details are omitted for brevity.

We can now prove the following theorem.

\begin{table*}[t]
\scriptsize \centering \caption{Minimum/Stopping Distance Results for Array LDPC Codes for Different Values of $q$ and $m$}\label{table:arrayLDPC}
\def\Hline{\noalign{\hrule height 2\arrayrulewidth}}
\vskip -3.0ex 
\begin{tabular}{cllllllll}
\Hline \\ [-2.0ex]
   $q$ & $h(q,7)$ & $d(q,7)$ & $h(q,6)$ & $d(q,6)$ & $h(q,5)$ & $d(q,5)$ & $h(q,4)$ & $d(q,4)$\\
\hline
\\ [-2.0ex] \hline  \\ [-2.0ex]
7 & $\mathbf{12}$ & $\mathbf{14}$ & $\mathbf{10}$ & $12$ \cite{sug08} & $\mathbf{9}$ & $12$ \cite{sug08} & 8 \cite{esm11} & 8 \cite{sug08}\\
11 &  $\mathbf{15}$ & $\mathbf{20}$ & $\mathbf{12}$ & $16$ \cite{sug08} & $10$ \cite{esm11} & $10$ \cite{sug08} & 10 \cite{esm11} & 10 \cite{sug08}\\
13 &  $\mathbf{16}$ & $\mathbf{20}$ &  $\mathbf{14}$ & $14$ \cite{sug08} & $\mathbf{12}$ & $12$ \cite{sug08} & $\mathbf{10}$ & 10 \cite{sug08}\\
17 & $\mathbf{18-24}$ & $\mathbf{18-24}$, even  & $\mathbf{16}$ & $\mathbf{16}$ & $\mathbf{12}$ & $12$ \cite{sug08} & $\mathbf{10}$ & 10 \cite{sug08}\\
19 & $\mathbf{18-20}$ & $\mathbf{18}$ or $\mathbf{20}$ & $\mathbf{16}$ & {\bf $\mathbf{18}$} & $\mathbf{12}$ & $12$ \cite{sug08} & $\mathbf{10}$ & 10 \cite{sug08}\\
23 & $\mathbf{17-22}$ & $\mathbf{18-22}$, even & $\mathbf{17-20}$ & $\mathbf{18}$ or $\mathbf{20}$& $\mathbf{12}$ & $\mathbf{12}$ & $\mathbf{10}$ & 10 \cite{sug08}\\
29 & $\mathbf{17-24}$ & $\mathbf{18-24}$, even & $\mathbf{17-20}$ & $\mathbf{18}$ or $\mathbf{20}$& $\mathbf{12}$ & $\mathbf{12}$ & $\mathbf{10}$ & 10 \cite{sug08}\\
31 & $\mathbf{17-24}$ & $\mathbf{18-24}$, even & $\mathbf{17-20}$ & $\mathbf{18}$ or $\mathbf{20}$& $\mathbf{12}$ & $\mathbf{12}$ & $\mathbf{10}$ & 10 \cite{sug08}\\
37 & $\mathbf{17-24}$ & $\mathbf{18-24}$, even & $\mathbf{17-20}$ & $\mathbf{18}$ or $\mathbf{20}$& $\mathbf{12}$ & $\mathbf{12}$ & $\mathbf{10}$ & 10 \cite{sug08}\\
41 & $\mathbf{17-24}$ & $\mathbf{18-24}$, even & $\mathbf{17-20}$ & $\mathbf{18}$ or $\mathbf{20}$& $\mathbf{12}$ & $\mathbf{12}$ & $\mathbf{10}$ & 10 \cite{sug08}\\
43 & $\mathbf{17-24}$ & $\mathbf{18-24}$, even & $\mathbf{17-20}$ & $\mathbf{18}$ or $\mathbf{20}$& $\mathbf{12}$ & $\mathbf{12}$ & $\mathbf{10}$ & 10 \cite{sug08}\\
47 & $\mathbf{17-24}$ & $\mathbf{18-24}$, even & $\mathbf{17-20}$ & $\mathbf{18}$ or $\mathbf{20}$& $\mathbf{12}$ & $\mathbf{12}$ & $\mathbf{10}$ & 10 \cite{sug08}\\
53 & $\mathbf{17-24}$ & $\mathbf{18-24}$, even & $\mathbf{17-20}$ & $\mathbf{18}$ or $\mathbf{20}$& $\mathbf{12}$ & $\mathbf{12}$ & $\mathbf{10}$ & 10 \cite{sug08}\\
59 & $\mathbf{17-24}$ & $\mathbf{18-24}$, even & $\mathbf{17-20}$ & $\mathbf{18}$ or $\mathbf{20}$& $\mathbf{12}$ & $\mathbf{12}$ & $\mathbf{10}$ & 10 \cite{sug08}\\
61 & $\mathbf{17-24}$ & $\mathbf{18-24}$, even & $\mathbf{17-20}$ & $\mathbf{18}$ or $\mathbf{20}$& $\mathbf{12}$ & $\mathbf{12}$ & $\mathbf{10}$ & 10 \cite{sug08}\\
67 & $\mathbf{17-24}$ & $\mathbf{18-24}$, even & $\mathbf{17-20}$ & $\mathbf{18}$ or $\mathbf{20}$& $\mathbf{12}$ & $\mathbf{12}$ & $\mathbf{10}$ & 10 \cite{sug08}\\
71 & $\mathbf{17-24}$ & $\mathbf{18-24}$, even & $\mathbf{17-20}$ & $\mathbf{18}$ or $\mathbf{20}$& $\mathbf{12}$ & $\mathbf{12}$ & $\mathbf{10}$ & 10 \cite{sug08}\\
73 & $\mathbf{17-24}$ & $\mathbf{18-24}$, even & $\mathbf{17-20}$ & $\mathbf{18}$ or $\mathbf{20}$& $\mathbf{12}$ & $\mathbf{12}$ & $\mathbf{10}$ & 10 \cite{sug08}\\
79 & $\mathbf{17-24}$ & $\mathbf{18-24}$, even & $\mathbf{17-20}$ & $\mathbf{18}$ or $\mathbf{20}$& $\mathbf{12}$ & $\mathbf{12}$ & $\mathbf{10}$ & 10 \cite{sug08} 
\end{tabular}
\end{table*}

\begin{theorem} \label{th:2}
The minimum distance $d(q,7)$ is upper-bounded by $24$ for $q > 7$.
\end{theorem}

\begin{IEEEproof}
  The proof is based on the candidate template support matrix shown in (\ref{eq:supportm7}) at the top of the page and is almost identical to the proof of Theorem~\ref{th:1}. In particular, it is easy to verify  that each entry in each row of the matrix appears an even number of times and that all columns in the matrix are in fact valid columns in an array LDPC code parity-check matrix (all columns are of the form in (\ref{eq:form})). 

For the third part of the proof, we need to show, for any value of $q > 7$, that there exist (at least two) columns in the candidate template support matrix which are not identical modulo $q$ and appear an odd number of times. Again, this is simple  (and very fast) to verify by a computer search for any finite value of $q$ that would be of any practical value. It is only for large values of $q$ that the theoretical proof below is needed. 

Now, let the largest absolute value of the entries in the $i$th row of the matrix in (\ref{eq:supportm7}) which do not involve a multiplication by $2^{-1}$ or $2^{-2}$ be denoted $\lambda_i$, and let the largest absolute value of the factor in front of $2^{-1}$ of the remaining entries in the $i$th row be denoted by $\mu_i$. Since $a \cdot 2^{-1} \pmod{q}$, when $a$ is odd (which is always the case in  (\ref{eq:supportm7})), can be written as $(q+a)/2$, it follows easily that for a row $i$ where all entries are of the form $a$ or $a \cdot 2^{-1}$, different template entries can never be the same modulo $q$ when $q > 2\lambda_i+\mu_i$. For the first row this bound is $2 \cdot 5 + 11 = 21$, and for the third row, this bound is $2 \cdot 2 + 5 = 9$. Thus, looking at the first row, the only possibility for repeated columns,  when $q > 21$ (the bound for the first row), is for two \emph{neighboring} columns (with identical entries in the first row) to be the same. 
However, by looking at the third row in the matrix, this possibility can be ruled out by requiring that $q > 9$ (the bound for the third row). 
In summary, it follows that there are no identical columns in the matrix in (\ref{eq:supportm7})  if $q > \max(21,9)=21$. 
Furthermore, for values of $7 < q < 21$, it can be verified numerically that there are no repeated columns in (\ref{eq:supportm7}), and the result follows. 
%
%
%
\end{IEEEproof}
As a final remark, for $q=7$, every column in the matrix in (\ref{eq:supportm7}) is repeated exactly twice, and the column-reduced version (as defined in Section~\ref{sec:formal}) will be the empty matrix. 



\section{Numerical Results} \label{sec:results}

In addition to the analytic results of Theorems~\ref{th:1} and \ref{th:2}, we have performed a computer search to compute the exact values for $d(q,m)$ and $h(q,m)$ for small values of $q$ and $m$. The results are summarized in Table~\ref{table:arrayLDPC}, where the entries that appear in bold are new results. Results from the literature have also been included with an explicit reference. 

For $m=6$, we have computed the exact values of $d(q,m)$ and $h(q,m)$ for $q \leq 19$. For larger values of $q$, we have run the exhaustive algorithm from \cite{ros09,ros12} with an upper weight/size threshold of $16$ without finding any codewords or stopping sets. From the upper bound of Theorem~\ref{th:1} and the fact that these codes are even-weight codes, we can conclude that  the minimum distance, for $23 \leq q \leq 79$, is either $18$ or $20$. Furthermore, extensive minimum distance calculations using the probabilistic algorithm from \cite{tom07} for several values of $q \geq 23$, indicate that the minimum distance is indeed $20$ for $q \geq 23$, from which it follows that  the upper bound from Theorem~\ref{th:1} appears to be tight.

For $m=7$, we have been able to compute the exact values of $d(q,m)$  and $h(q,m)$ for $q=7$, $11$, and $13$.  
For $q=13$, we were able to run the exhaustive algorithm from \cite{ros09,ros12} with an upper weight threshold of $18$ without finding any codewords.  In addition, we found a codeword of weight $20$ using the probabilistic algorithm from \cite{tom07}, from which (and the fact that the array LDPC codes are even-weight codes) we can conclude that the minimum distance is  indeed $20$. 
For larger values of $q$, $17 \leq q \leq 29$, the probabilistic algorithm from \cite{tom07} has provided the upper bounds in Table~\ref{table:arrayLDPC}. Note that even if the results are formally stated as upper bounds, the algorithm from \cite{tom07} indicates that the upper bounds  are indeed likely to give the exact values, which again indicates that the bound from Theorem~\ref{th:2} is in fact tight (for instance, $q=17$ gives a minimum distance of $24$ with very high probability). 
For the high values of $q$ ($31 \leq q \leq 79$), Theorem~\ref{th:2} has provided the upper bounds. The lower bounds on $d(q,7)$ and $h(q,7)$, for $q \geq 17$, have been established by running the exhaustive algorithm from \cite{ros09,ros12} with an upper weight/size threshold of $16/17$ for $q=17$ and $19$ and an upper weight/size threshold of $16$ for $q \geq 23$ without finding any codewords or stopping sets. 



\section{Conclusion and Future Work} \label{sec:conclu}
In this paper, the minimum/stopping distance of array LDPC codes has been studied. We have presented an improved general (i.e., independent of $q$) upper bound on the minimum distance for the case $m=6$, using the concept of a template support matrix of a codeword/stopping set, which significantly improves the currently best known bound.   The bound appears to be tight with high probability in the sense that we have not found  codewords of strictly lower weight  for several values of $q$ using a minimum distance probabilistic algorithm.  In addition, we have provided the new upper bound $d(q,7) \leq 24$ which also (from extensive numerical computations) appears to be tight. Finally, we have provided several new specific minimum/stopping distance results for $m \leq 7$ and low-to-moderate values of $q \leq 79$.

We believe that extending the approach of this paper to larger values of $m$ is an important topic for future work. Currently, the main bottleneck is to find a sufficient number of low-weight/small-size codewords/stopping sets when $m$ grows (and $q$ is not too large), since current state-of-the-art algorithms for finding low-weight/small-size codewords/stopping sets fail in such scenarios. Another important question for future work would be to determine whether or not it is always possible to find a template support matrix for any fixed value of $m$, which would imply that the minimum/stopping distance is upper-bounded by a constant (depending only on $m$) for any fixed value of $m$.

\section*{Acknowledgment}

The authors wish to thank the Associate Editor Prof.\ Burshtein and the anonymous reviewer for their valuable comments and suggestions that helped improve the presentation of the paper.

\balance



\begin{thebibliography}{10}
\providecommand{\url}[1]{#1}
\csname url@samestyle\endcsname
\providecommand{\newblock}{\relax}
\providecommand{\bibinfo}[2]{#2}
\providecommand{\BIBentrySTDinterwordspacing}{\spaceskip=0pt\relax}
\providecommand{\BIBentryALTinterwordstretchfactor}{4}
\providecommand{\BIBentryALTinterwordspacing}{\spaceskip=\fontdimen2\font plus
\BIBentryALTinterwordstretchfactor\fontdimen3\font minus
  \fontdimen4\font\relax}
\providecommand{\BIBforeignlanguage}[2]{{%
\expandafter\ifx\csname l@#1\endcsname\relax
\typeout{** WARNING: IEEEtran.bst: No hyphenation pattern has been}%
\typeout{** loaded for the language `#1'. Using the pattern for}%
\typeout{** the default language instead.}%
\else
\language=\csname l@#1\endcsname
\fi
#2}}
\providecommand{\BIBdecl}{\relax}
\BIBdecl

\bibitem{fan00}
J.~L. Fan, ``Array codes as low-density parity-check codes,'' in \emph{Proc.
  2nd Int. Symp. Turbo Codes {\&} Rel. Topics}, Brest, France, Sep. 2000, pp.
  543--546.

\bibitem{mit02}
T.~Mittelholzer, ``Efficient encoding and minimum distance bounds of
  {R}eed-{S}olomon-type array codes,'' in \emph{Proc.~IEEE
  Int.~Symp.~Inf.~Theory (ISIT)}, Lausanne, Switzerland, Jun./Jul. 2002, p.
  282.

\bibitem{yan03}
K.~Yang and T.~Helleseth, ``On the minimum distance of array codes as {LDPC}
  codes,'' \emph{IEEE Trans.~Inf.~Theory}, vol.~49, no.~12, pp. 3268--3271,
  Dec. 2003.

\bibitem{sug08}
K.~Sugiyama and Y.~Kaji, ``On the minimum weight of simple full-length array
  {LDPC} codes,'' \emph{IEICE Trans.\ Fundamentals}, vol. E91-A, no.~6, pp.
  1502--1508, Jun. 2008.

\bibitem{esm09}
M.~Esmaeili and M.~J. Amoshahy, ``On the stopping distance of array code
  parity-check matrices,'' \emph{IEEE Trans.~Inf.~Theory}, vol.~55, no.~8, pp.
  3488--3493, Aug. 2009.

\bibitem{esm11}
M.~Esmaeili, M.~H. Tadayon, and T.~A. Gulliver, ``More on the stopping and
  minimum distances of array codes,'' \emph{IEEE Trans.~Commun.}, vol.~59,
  no.~3, pp. 750--757, Mar. 2011.

\bibitem{liu10}
H.~Liu, L.~Ma, and J.~Chen, ``On the number of minimum stopping sets and
  minimum codewords of array {LDPC} codes,'' \emph{IEEE Commun.~Lett.},
  vol.~14, no.~7, pp. 670--672, Jul. 2010.

\bibitem{dol10}
L.~Dolecek, Z.~Zhang, V.~Anantharam, M.~J. Wainwright, and B.~Nikolic,
  ``Analysis of absorbing sets and fully absorbing sets of array-based {LDPC}
  codes,'' \emph{IEEE Trans.~Inf.~Theory}, vol.~56, no.~1, pp. 181--201, Jan.
  2010.

\bibitem{olc03}
S.~\"Ol\c{c}er, ``Decoder architecture for array-code-based {LDPC} codes,'' in
  \emph{Proc.\ IEEE Global Telecommun. Conf. (GLOBECOM)}, vol.~4, San
  Francisco, CA, Dec. 2003, pp. 2046--2050.

\bibitem{bha05}
P.~Bhagawat, M.~Uppal, and G.~Choi, ``{FPGA} based implementation of decoder
  for array low-density parity-check codes,'' in \emph{Proc.\ IEEE Int. Conf.
  Acoustics, Speech, and Signal Processing (ICASSP)}, vol.~5, Philadelphia, PA,
  Mar. 2005, pp. 29--32.

\bibitem{ros09}
E.~Rosnes and {\O}.~Ytrehus, ``An efficient algorithm to find all small-size
  stopping sets of low-density parity-check matrices,'' \emph{IEEE
  Trans.~Inf.~Theory}, vol.~55, no.~9, pp. 4167--4178, Sep. 2009.

\bibitem{ros12}
E.~Rosnes, {\O}.~Ytrehus, M.~A. Ambroze, and M.~Tomlinson, ``Addendum to ``{An}
  efficient algorithm to find all small-size stopping sets of low-density
  parity-check matrices'','' \emph{IEEE Trans.~Inf.~Theory}, vol.~58, no.~1, pp.
  164--171, Jan. 2012.

\bibitem{tom07}
M.~Tomlinson, C.~Tjhai, J.~Cai, and M.~Ambroze, ``Analysis of the
  distribution of the number of erasures correctable by a binary linear code
  and the link to low-weight codewords,'' \emph{IET Commun.}, vol.~1, no.~3,
  pp. 539--548, Jun. 2007.

\end{thebibliography}

\end{document}